\documentclass[12pt,preprint]{article}
\usepackage{amssymb}
\usepackage{amsmath}
\usepackage{graphicx}

\graphicspath{{img/}}

\begin{document}

\begin{titlepage}
  \vspace{0.6cm}

  \begin{center}
    \Large{Modification of black-body radiance at low temperatures and frequencies}
    \vspace{1.0cm}

    \large{Carlos Falquez$^\dagger$, Ralf Hofmann$^*$, and Tilo Baumbach$^\dagger$}
  \end{center}
  \vspace{1.0cm}

\begin{center}
    {\em $\mbox{}^\dagger$
    Laboratorium f\"ur Applikationen der Synchrotronstrahlung (LAS)\\
    Karlsruher Institut f\"ur Technologie (KIT)\\
    Postfach 6980\\
    D-76128 Karlsruhe\\
    Germany}\\
  \end{center}
  \vspace{1.5cm}
  \begin{center}
    {\em $\mbox{}^*$ Institut f\"ur Theoretische Physik\\
    Universit\"at Heidelberg\\
    Philosophenweg 16\\
    69120 Heidelberg, Germany}
  \end{center}
  \vspace{1.0cm}

  \begin{abstract}
    In contrast to earlier reports, where the spectrum of the {\sl energy density} of photonic black-body
    radiation modified by SU(2) effects was discussed, we discuss the low-frequency spectrum of the
    {\sl radiance} at temperatures ranging between 5 and 20 Kelvin. 
We conclude that compared to the conventional theory the only observable 
    effect is associated with the spectral gap (total screening) which for $T\ge 4.3\,$K scales 
with temperature $T$ as $\frac{\nu^*(T)}{\text{GHz}}=42.70 \left( \frac{T}{\text{K}} \right)^{-0.53} + 0.21\,$. We also 
discuss how a low-temperature 
    black body cavity under the influence of a sufficiently strong 
static electric field is forced to emit according to Planck's radiation law (pure U(1) theory) 
even at low frequencies and how this effect can be utilized to measure SU(2) induced deviations. 
  \end{abstract}
\end{titlepage}

\section{Introduction}

The possibility that the U(1) symmetry underlying the propagation of photons as described by 
the Standard Model of particle physics is the result of a dynamical gauge symmetry breaking of an 
SU(2) Yang-Mills theory is theoretically \cite{Hofmann2005} and observationally 
intriguing \cite{PSA2005,JHEP2007,JHEP2008,CD2008}. In a thermal version of pure and deconfining 
SU(2) gauge theory associated with a single temperature $T$ 
this symmetry breaking is induced by the effective thermal ground state which emerges 
upon a spatial coarse-graining over interacting calorons of topological charge modulus 
one \cite{HH2004,GH2006}. The relevant calorons (of trivial holonomy) are stable, periodic, minimal-action, zero energy-momentum, and 
thus nonpropagating solutions to the Euclidean Yang-Mills equation which possess unit topological-charge modulus \cite{HS1977}. 
The emerging thermal ground state provides for the temperature-dependent massiveness of two out of 
the three propagating gauge modes while a third direction of the SU(2) algebra (the photon) remains 
massless. The fact that slightly below the critical temperature $T_c$ a tiny photon mass 
due to the Meissner effect induced by a condensate 
of (electric) monopoles seems to be implicit in the low-frequency CMB data, see \cite{Arcade2} and references therein 
and \cite{RH2009} for an interpretation, fixes $T_c$ to be very closely above the present CMB baseline 
temperature of about 2.73\,Kelvin. 

In \cite{RH2006,SHG2007,KH2007,LH2008} an account of the 
computation of radiative corrections in the effective theory has 
been given. In particular, the screening function $G$ for thermalized photon propagation in the deconfining phase was 
computed exactly on the one-loop level\footnote{This is more than sufficient for any practical purpose, 
see \cite{RHLeip2007}.} in \cite{LH2008}. Our predictions for the modified black-body spectrum that rely on function $G$ so 
far were for the spectral energy density (usually referred to as `spectral intensity' in 
\cite{JHEP2008,SHG2007,LH2008}) which is not directly accessible
experimentally\footnote{Academically, it is in principle accessible through gravitational
  interactions.}. For bolometry and radiometry the relevant quantity is the spectral 
radiance $L$.  

The purpose of this note is an investigation of the characteristics of 
the radiance spectrum at low temperature and frequency as well as a discussion of 
observable effects induced by SU(2) gauge dynamics. 

\section{Bolometry and radiometry of SU(2) photons}

Here we derive the photonic radiance of an SU(2) Yang-Mills theory subject to
a modified dispersion relation at low temperatures and frequencies \cite{LH2008} 
in comparison with the conventional U(1) Planck spectrum. In contrast to
earlier publications, where natural units were used \cite{JHEP2008,SHG2007,LH2008},
we exclusively work in SI units in this report.
Recall that the spectral energy density $I(\omega)$ of a photon gas in thermal equilibrium is given by
the number of modes per volume available within a frequency interval times the average (thermal)
energy per mode. One has 
\begin{equation}
\label{energydens}
  \begin{split}
    dE&=I(\omega)d\omega=2 \frac{d^3 \mathbf{k}}{\left(2 \pi \right)^3} \times \frac{\hbar \omega}{e^{\frac{\hbar \omega}{kT}}-1}\\
      &=\frac{4}{h^2}\frac{\omega}{e^{\frac{\hbar \omega}{kT}}-1} \mathbf{p}^2(\omega)  \frac{dp}{d\omega} d\omega\\
      \Rightarrow I(\nu)&=\frac{8 \pi}{h^2}\frac{\omega}{e^{\frac{\hbar \omega}{kT}}-1} \mathbf{p}^2(\omega)\,\frac{dp}{d\omega}\,,
  \end{split}
\end{equation}
where $k$ is Boltzmann's constant, $h$ is Planck's quantum of action
($\hbar=\frac{h}{2\pi}$), $\mathbf{p}$ denotes a photon's spatial momentum, 
$\omega=2\pi\nu$, and $T$ is the absolute temperature.
The factor $\frac{dp}{d\omega}$ in Eq.\,(\ref{energydens}) accounts for the dispersion
relation $\omega=\omega(|\mathbf{p}|)$. In a U(1) theory this relation is given by
$(\hbar\omega)^2=(cp_0)^2=c^2\mathbf{p}^2$ with $c$ the speed of light in vacuum.
For the photons of an SU(2) Yang-Mills theory the dispersion law is
substantially altered at low frequencies and low temperatures, an effect which is
described by the (anti)screening function\footnote{Microscopically, this
  screening or antiscreening takes place by monopole-antimonopole creation or 
successive scattering of photons off monopoles and antimonopoles,
respectively.} $G$. 
Namely, one has 
\begin{equation}
\label{moddsip}
p_0^2-\mathbf{p}^2=G(|\mathbf{p}|,T)\,.
\end{equation}
For photons to propagate it must hold that $p_0^2\ge G$ (otherwise $|\mathbf{p}|$ becomes imaginary).
This restricts the possible values of $\omega$ associated with energy transport 
to $\omega \ge \omega^*$, with $\omega^*$ defined as the root of $|\mathbf{p}(\omega^*)|=0$.

In conventional U(1) theory both quantities, spectral energy density $I(\nu)$
and spectral radiance $L(\nu)$, are proportional to one another,
$L(\nu)=\frac{c}{4\pi}\times I(\nu)$ \cite{GrumBech79}.
In a deconfining SU(2) plasma, however, $c$ must be replaced by the photon's group velocity
$v_g$ defined as
\begin{equation}
\label{groupvel}
v_g\equiv
\partial_{|\mathbf{p}|}\,E=\hbar\,\partial_{|\mathbf{p}|}\,\omega=\partial_{|\mathbf{k}|}\,\omega\,,
\end{equation}
where $E=h\nu=cp_0$ is photonic energy and $\mathbf{k} \equiv \mathbf{p} / \hbar$ the wave number vector.
Thus in calculating the spectral radiance $L(\nu)$ for SU(2) photons, the
factor $\frac{dp}{d\omega}$ cancels the one in Eq.\,(\ref{energydens}), and we obtain
\begin{equation}
    \begin{split}
      L^{\tiny\mbox{SU(2)}}_\nu \left( T,\nu \right) &= \frac{2 h}{c^2} \frac{\nu^3}{e^{\frac{h \nu}{k T} } -1 } \times
                \left(1 - \frac{c^2 G}{(h\nu)^2}\right)
                \theta \left( \nu - \nu^* \right)\\
                &= L^{\tiny\mbox{U(1)}}_\nu \times
                   \left(1 - \frac{c^2 G}{(h\nu)^2}\right) \theta \left( \nu-\nu^* \right)
     \end{split}
     \label{LSU2}
\end{equation}
where $L^{\tiny\mbox{U(1)}}_\nu$ denotes the Planckian spectral radiance.

As shown in \cite{LH2008}, the (anti)screening function $G$ may be calculated 
in a selfconsistent way using 
numerical methods. To make contact with the real world the critical
temperature $T_c$ for the deconfining-preconfining phase transition, 
which is the only free parameter of a thermalized SU(2) 
quantum Yang-Mills theory, must be determined experimentally.
In \cite{RH2009} we have given observational reasons why $T_c$ should be very closely above 
the baseline temperature of about 2.73\,Kelvin of the present cosmic microwave background.  

\begin{figure}[htpb]
  \centering
  \includegraphics[width=120mm]{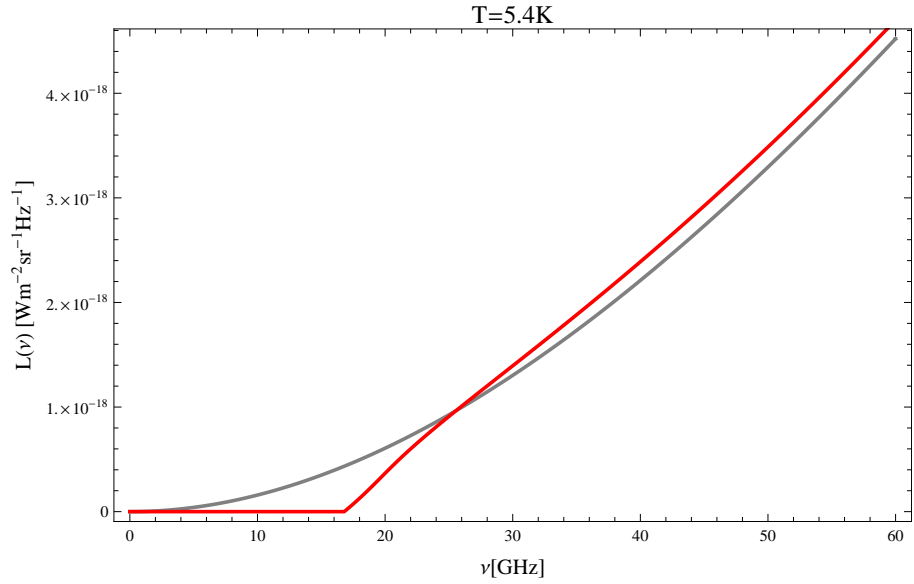}
  \caption{Comparison between the black body spectral radiances for $T=5.4$\,K of
    an SU(2) (red) and a U(1) (grey) theory.}
  \label{054B}
\end{figure}

\begin{figure}[htpb]
  \centering
  \includegraphics[width=120mm]{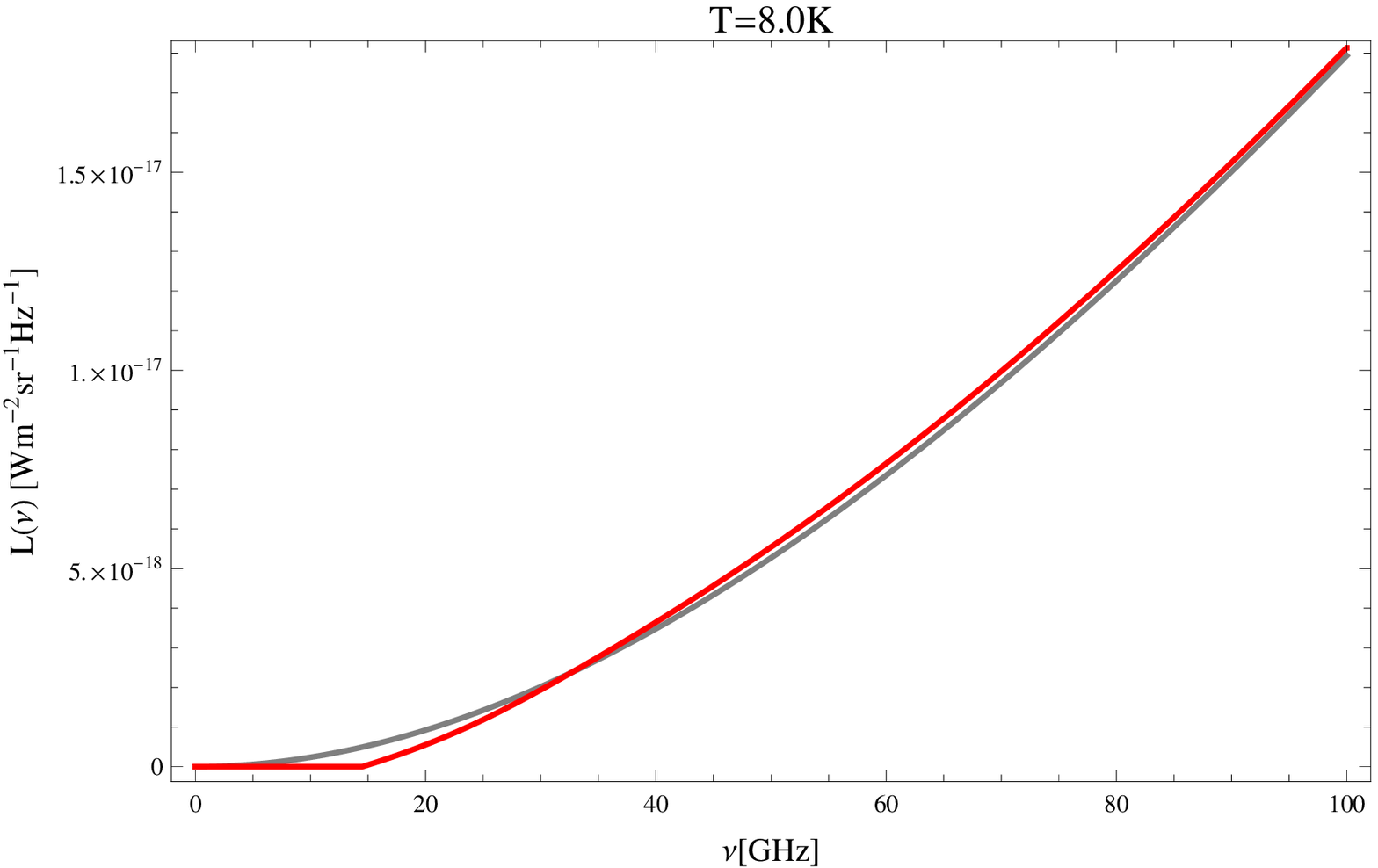}
  \caption{Comparison between black body spectral radiances for $T=8.0$\,K of
    an SU(2) (red) and a U(1) (grey) theory.}
  \label{080B}
\end{figure}

\begin{figure}[htpb]
  \centering
  \includegraphics[width=120mm]{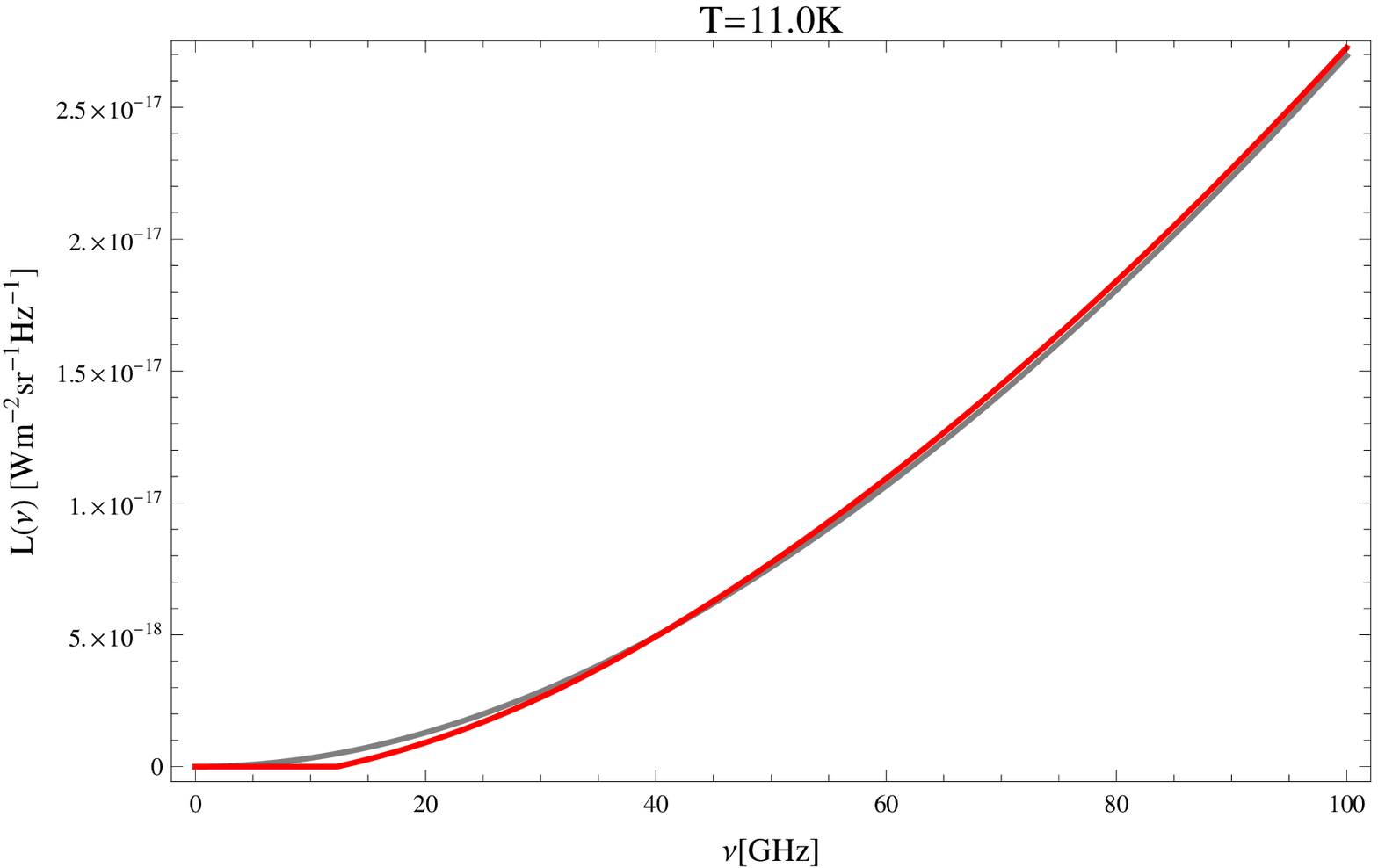}
  \caption{Comparison between black body spectral radiances for $T=11.0$\,K of
    an SU(2) (red) and a U(1) (grey) theory.}
  \label{110B}
\end{figure}

\begin{figure}[htpb]
  \centering
  \includegraphics[width=120mm]{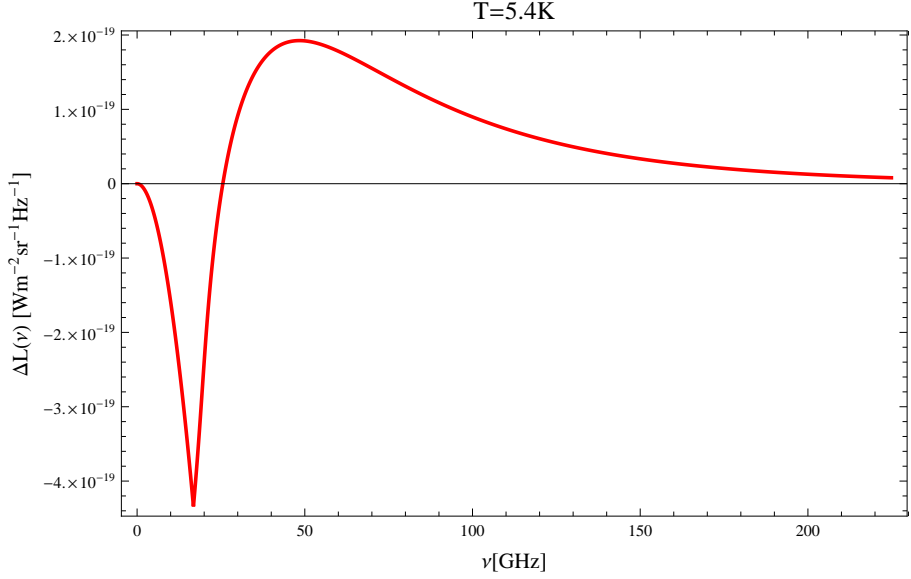}
  \caption{The difference in spectral radiance 
$\Delta L(\nu,T)$ at $T=5.4$\,K as a
    function of frequency $\nu$.}
  \label{054dL}
\end{figure}

After fixing $T_c$ the modified radiance spectra for SU(2) photons may be calculated for various 
physical temperatures. In Figs.\,\ref{054B}, \ref{080B}, \ref{110B} we show results for
$T=5.4, 8.0,$ and 11.0 Kelvin, respectively. The red lines correspond to the
calculated SU(2) radiance, the conventional U(1) Planck spectrum is depicted in grey.
Notice the regime of total screening (suppression of spectral radiance down to zero)
and the cross-over to a regime of slight antiscreening (excess of spectral radiance)
in all cases. Fig.\,\ref{054dL} shows the difference in spectral radiance
$\Delta L_\nu\left( T,\nu \right)$, defined as
\begin{equation}
\Delta L_\nu\left(T,\nu \right) = L^{\tiny\mbox{SU(2)}}_\nu\left(T,\nu \right)-L^{\tiny\mbox{U(1)}}_\nu\left(T,\nu \right)\,
\label{deltaL}
\end{equation}
for $T=5.4$\,K. 

A type of bolometric experiment can be conceived as follows. Let the apertures of
an isolated low-temperature U(1) black body at temperature $T_1$ and
that of an SU(2) black body of identical characteristics at temperature $T_2$ face each other,
and exchange radiant energy. Have the SU(2) black body be linked to a large heat reservoir to
keep its wall temperature $T_2$ constant. Furthemore, switch in a band-width 
filter within the common aperture tuned to the region of the SU(2) black-body gap: Such a filter absorbs photons above 
frequency $\nu^*$ no matter which cavity they come from\footnote{
From the three effective SU(2) gauge modes only the tree-level massless mode (the photon) interacts with electric 
charges and so can be absorbed by the material in the band-width filter. If a propagating photon emerges from the SU(2) 
cavity then the absorbed power per frequency interval is identical to that of a propagating photon stemming from the U(1) 
cavity because the additional factor $\frac{1}{1-\frac{c^2 G}{(h\nu)^2}}$ in the SU(2) 
spectral radiance, see Eq.\,(\ref{LSU2}), is canceled, see discussion below Eq.\,(\ref{spectrpowerapp}).}.  
U(1) photons within the SU(2) spectral gap are absorbed by the SU(2) thermal ground 
state \cite{LudescherEtAl2008}: They create unresolvable
monopole-antimonopole pairs. Thus a small amount of energy per time flows across the common aperture 
from the U(1) cavity to the SU(2) cavity, and zero radiation temperature within the U(1) cavity will asymptotically 
be reached (radiation refrigeration).   

For $T_1=T_2$ and blocking off the gap region $0\le\nu\le\nu^*(T_1)$ by a complementary band-width filter, 
the regime of propagating photon frequencies would not allow for any heat exchange since   
 $L^{\tiny\mbox{SU(2)}}(\nu,T_2)\,d\Omega_{\tiny\mbox{SU(2)}}$ and 
$L^{\tiny\mbox{U(1)}}(\nu,T_1=T_2)\,d\Omega_{\tiny\mbox{U(1)}}$ do match. 
Here $d\Omega_{\tiny\mbox{SU(2)}}$ and 
$d\Omega_{\tiny\mbox{U(1)}}$ are solid-angle elements defined on a sphere centered at a point within the common aperture. 
By Snell's law \cite{krall} we have 
\begin{equation}
\label{snell's law}
\frac{d\Omega_2}{d\Omega_1}=\frac{v^2_{ph}}{c^2}\,,
\end{equation}
where the phase velocity $v_{ph}$ of energy propagation inside the SU(2) blackbody is given by
$v_{ph}\equiv\frac{\omega}{k}=\frac{c}{\sqrt{1-\frac{c^2 G}{(h\nu)^2}}}$, and therefore 
\begin{equation}
\label{matchL}
 L^{\tiny\mbox{SU(2)}}(\nu,T_2)\,\frac{d\Omega_{\tiny\mbox{SU(2)}}}{d\Omega_{\tiny\mbox{U(1)}}}=
L^{\tiny\mbox{U(1)}}(\nu,T_1=T_2)\,.
\end{equation}
A potentially interesting quantity for experiments is the 
\emph{radiance U(1) line temperature} $T_P(\nu)$. The quantity $T_P(\nu)$ is
the temperature a conventional U(1) black body must possess 
in order to reach the following (bolometric) equilibrium condition \cite{krall}, compare
with Eq.\,(\ref{snell's law}):
\begin{equation}
\label{equilcond}
L^{U(1)}_\nu=L^{SU(2)}_\nu \times \frac{1}{1-\frac{c^2 G}{(h\nu)^2}}\,.
\end{equation}
One has
\begin{equation}
\label{linetempdef}
  T_P \left(L^{\tiny\mbox{SU(2)}}_\nu(\nu,T)\right)\equiv
\frac{h \nu}{k} \frac{1}{\ln \left[ \frac{2h}{c^2}\,\left(1-\frac{c^2 G}{(h\nu)^2}\right)\,\frac{\nu^3}{L^{\tiny\mbox{SU(2)}}
_\nu(\nu,T)} + 1 \right]}\,.
\end{equation}
Again, on the right-hand side of Eq.\,(\ref{equilcond}) the factor $\left(1-\frac{c^2 G}{(h\nu)^2}\right)$,
which apart from the factor $\theta \left( \nu - \nu^* \right)$ distinguishes SU(2) from U(1) spectral radiance, 
see Eq.\,(\ref{LSU2}), is canceled. Therefore $T_P$ is the same as the SU(2) wall temperature $T_2$ except for the 
regime of total screening where $T_P\equiv 0$. That is, no heat is effectively exchanged by 
frequencies above the spectral gap according to the bolometric equilibrium condition  (\ref{equilcond}), 
and within the spectral gap no U(1) photons are required in the balance of Eq.\,(\ref{linetempdef}). 
An essential question, addressed in Sec.\,\ref{sec:prepU1}, is how the above-assumed hypothetic 
low-temperature U(1) black body can be realized experimentally. 

\begin{figure}[htpb]
  \centering
  \includegraphics[width=120mm]{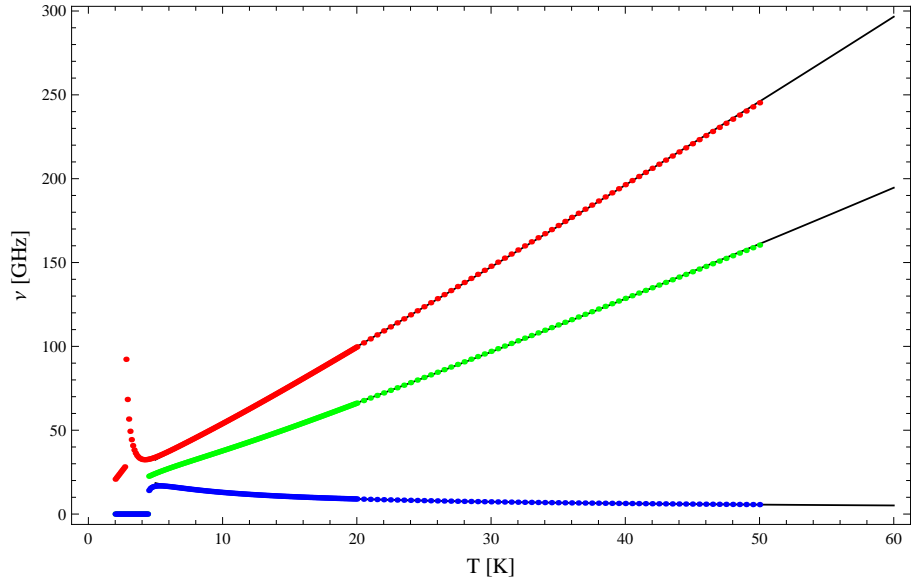}
  \caption{Plots of the $T$ dependence of the characteristic points of $L^{SU(2)}_\nu(T,\nu)$. 
The blue points depict $\nu^*(T)$,
  the green points $\nu_c(T)$, and the red points $\nu_M(T)$. }
  \label{Ypoints}
\end{figure}

For completeness let us give some characterization of the factor 
\begin{equation*}
\left(1 - \frac{c^2 G}{(h\nu)^2}\right)\theta \left( \nu - \nu^* \right)
\end{equation*}
 which converts U(1) to SU(2) spectral radiance and comprises of 
the characteristic frequencies $\nu^*$, $\nu_c$, $\nu_M$ which are implicitely defined as follows:
\begin{equation}
  \begin{split}
    |\mathbf{p}|(\nu^*) &= 0\,,\\
    G(\nu_c,T) &= 0\,,\ \ \mbox{and}\\
    \frac{G(\nu_M,T)}{\nu_M^2} &= \min\{\frac{G(\nu,T)}{\nu^2}\}\,.
  \end{split}
\end{equation}
Lowering $\nu$ at fixed $T$, the points $\nu^*$, $\nu_c$, and $\nu_M$ describe the onset of total 
screening (no photon propagation), the cross-over between screening and antiscreening ($G=0$), and the maximal antiscreening ($G<0$), respectively.
For $T>8\,$K the critical points $\nu_c$, $\nu_M$ and $\nu^*$ were numerically fitted to a power law in 
$T$. We obtain the following results: 
\begin{equation}
\label{paramser}
  \begin{split}
    \frac{\nu_c(T)}{\text{GHz}} &= 1.83 \left( \frac{T}{\text{K}} \right)^{1.12} + 13.48\\
    \frac{\nu_M(T)}{\text{GHz}} &= 3.45 \left( \frac{T}{\text{K}} \right)^{1.08} + 12.90\\
    \frac{\nu^*(T)}{\text{GHz}} &= 42.70 \left( \frac{T}{\text{K}} \right)^{-0.53} + 0.21\,.
  \end{split}
\end{equation}
Fig.\,\ref{Ypoints} shows calculated points overlaid with the fitted curves.

We now consider a radiometric approach by placing an antenna inside the SU(2)
plasma. Independent of (lossless) propagation properties, one arrives at the
following expression for the power $P_{\Delta\nu}(\nu)$ within band width
$\Delta\nu$ absorped by the antenna \cite{GrumBech79}: 
\begin{equation}
\label{spectrpower}
P_{\Delta\nu}(\nu)=\theta(\nu-\nu^*)\,\int_\nu^{\nu+\Delta\nu} d\nu^\prime\,\frac{h\nu^\prime}{\exp\left[\frac{h\nu^\prime}{kT}\right]-1}\,.
\end{equation}
For $h\nu\ll kT$, which is certainly the case for the spectral range we are
interested in, Eq.\,(\ref{spectrpower}) simplifies as
\begin{equation}
\label{spectrpowerapp}
P_{\Delta\nu}(\nu) \approx \theta(\nu-\nu^*)\,\Delta\nu kT\,.
\end{equation}
Apart from the $\theta$-function prefactor in Eqs.\,(\ref{spectrpower})
and (\ref{spectrpowerapp}) the expressions are identical to the U(1) situation
since a factor $|\mathbf{k}|^2$ in $L^{SU(2)}_\nu(\nu,T)$ is cancelled by a 
factor $|\mathbf{k}|^{-2}$, see \cite{GrumBech79}.
Radiometric measurements inside the predicted screening regime may therefore be able
to test for the presence of a fundamental SU(2) ground state resulting from gauge dynamics.

\section{Preparation of a U(1) black body at low temperatures}
\label{sec:prepU1}

The concept of a U(1) line temperature relies on one's ability
to prepare a black-body source in such a way that it emits according to Planck's
radiation law even at small temperatures where we expect modifications thereof. Since a homogeneous
electric field of energy density well above that of the thermal ground state
effectively switches off the indirect coupling of the photon to the ground state via scattering processes
involving the massive vector modes a black-body cavity with the bulk of
its volume transcended by such an electric field effectively acts as a U(1)
emitter. In the single-photon counting experiment of 
Ref.\,\cite{Tada2006} the (two-step laser) excitation of the 111$_{s_{1/2}}$ Rydberg state of $^{85}$Rb 
atoms towards the 111$_{p_{3/2}}$ state by absorption of thermal photons of 
frequency 2527\,MHz, prepared within a tunable cavity, can be exploited to learn about the temperature 
dependence of the mean photon number $\bar{n}(T)$ at this frequency. The measurement was 
performed for temperatures $T$ ranging from 67\,mK up to 1\,K. As discussed in \cite{Tada2006}, 
no deviation of $\bar{n}(T)$ from the U(1) expected Bose-Einstein distribution 
was observed. It is important to note that a static, electric stray field of 
$|\vec{E}|\sim 25\,$mV/cm was present in the cavity during the experiment. 

Such an electric field, however, would cause a sizable distortion of the thermal SU(2) 
ground state which is sufficient to render the system effectively U(1). 
The (unresolvable) electric monopoles residing in the thermal ground 
state are accelerated by the external field in a parallel or antiparallel way, thus 
acquire kinetic energy which they subsequently disperse by collisions thereby 
increasing the energy density of the thermal ground state 
to an effective temperature largely disparate to the temperature of the radiation which is kept in thermal equilibrium 
with the cavity walls. This separation of excitation from ground 
state physics frees the propagation of photons from any ground-state induced effect and 
thus renders their dispersion law trivial, that is, of the convential U(1) type. 
This effect may be used to test whether thermalized but otherwise unadulterated photon propagation that 
we observe in Nature really is a manifestation of strongly interacting thermal SU(2) gauge dynamics.   

The energy density $\rho_E$ of the external electric field is given as 
\begin{equation}
\label{enEfeld}
\rho_E=\frac{\epsilon_0}{2}\,\vec{E}^2\,,
\end{equation}
where in SI units $\epsilon_0=8.8542\,10^{-12}\,$J/(Vm$^2$). The energy density of the SU(2) thermal ground state 
$\rho_{gs}$ in SI units is given as
\begin{equation}
\label{gsthe}
\rho_{gs}=4\pi\,\Lambda_{CMB}^3 \frac{k_B}{(\hbar c)^3} T\,,
\end{equation}
where $\Lambda_{CMB}=2\pi\,\frac{k_B T_c}{\lambda_c}$, $T_c=2.725\,$K \cite{RH2009}, 
$\lambda_c=13.87$ \cite{Hofmann2005}, $k_B=1.3807\,10^{-23}\,$J/K, 
$c=2.9979\,10^8$m/s, $\hbar=\frac{6.6261}{2\pi}\,10^{-34}\,$Js. 
Setting $\rho_E=\rho_{gs}$ at $|\vec{E}|\sim 25\,$mV/cm \cite{Tada2006}, we obtain an effective 
ground-state temperature of t$10^3$\,K. Thus the energy density of the external electric 
field is more than three orders of magnitude larger than that of 
the SU(2) thermal ground at 1\,K and below. Thus we may safely assume a decoupling of the ground-state 
physics from the propagation properties of photons. In a rough estimate on an 
admissible field strength to not distort the SU(2) ground-state physics 
sizably $\rho_E$ should be less than $\rho_{gs}$. For example, demanding that 
$\rho_E\sim 0.1\,\rho_{gs}$ at $T=5.4\,$K implies an electric field 
strength of about $|\vec{E}|\sim 0.2\,$mV/cm. Thus to produce a U(1) black body at low temperature 
one should work with a much larger 
value of $|\vec{E}|$. A more refined treatment of the effects induced on the thermal ground state 
by external fields, considered as small perturbations to undistorted SU(2) Yang-Mills thermodynamics, 
should be performed within the realm of linear-reponse 
theory. We leave this to future investigation.

\section{Summary and Conclusions\label{SC}}

In this note we have investigated the experimental consequences of the 
assumption that, fundamentally, photon propagation is described by an SU(2)
rather than U(1) gauge principle. We have considered bolometric 
and radiometric methods, and we have shown that the only region where a differences to 
the conventional theory can be predicted is associated with the spectral gap $0\le\nu\le\nu^*(T)$ with the $T$ dependence of $\nu^*$ 
given in Eq.\,(\ref{paramser}) (total screening). We also have elucidated how a U(1) black body can be prepared at low temperature by 
virtue of decoupling its thermal ground state from its excitations due to the application of a static, homogeneous 
electric field. 

\section*{Acknowledgments}
CF and RH would like to thank Markus Schwarz for useful conversations.

\clearpage

\bibliographystyle{hieeetr}
\bibliography{Bibliography}

\end{document}